# Transmission Expansion Planning for Renewable-energy-dominated Power Grids Considering Climate Impact

Jin Lu, *Student Member, IEEE*, and Xingpeng Li, *Senior Member, IEEE*

*Abstract*—As renewable energy is becoming the major resource in future grids, the weather and climate can have a higher impact on grid reliability. Transmission expansion planning (TEP) has the potential to reinforce a transmission network that is suitable for climate-impacted grids. In this paper, we propose a systematic TEP procedure for climate-impacted renewable energy-enriched grids. Particularly, this work developed an improved model for TEP considering climate impact (TEP-CI) and evaluated the system reliability with the obtained transmission investment plan. Firstly, we created climate-impacted spatio-temporal future grid data to facilitate the TEP-CI study, which includes the future climate-dependent renewable production as well as the dynamic rating profiles of the Texas 123-bus backbone transmission system (TX-123BT). Secondly, we proposed the TEP-CI which considers the variation in renewable production and dynamic line rating, and obtained the investment plan for future TX-123BT. Thirdly, we presented a customized security-constrained unit commitment (SCUC) specifically for climate-impacted grids. The future grid reliability under various investment scenarios is analyzed based on the daily operation conditions from SCUC simulations. The whole procedure presented in this paper enables numerical studies on grid planning considering climate impacts. It can also serve as a benchmark for other TEP-CI research and performance evaluation.

<div style="text-align:center">NOMENCLATURE</div>

*Sets*

| | |
|---|---|
| $D^T$ | Set of typical days in a year. |
| $D^{T(D)}$ | Set of typical weekdays in a year. |
| $D^{T(E)}$ | Set of typical weekends in a year. |
| $G$ | Set of existing generators in the power system. |
| $G(p)$ | Set of online generators in period $p$. |
| $G'$ | Set of new generators in the power system. |
| $G^{B(n)}$ | Set of generators located on bus $n$. |
| $G^{N,B(n)}$ | Set of new generators located on bus $n$. |
| $G(p)$ | Set of generators including operational new generators in future period $p$. |
| $L'$ | Set of candidate new lines. |
| $L^{T(n)}$ | Set of transmission lines which's to bus is $n$. |
| $L^{F(n)}$ | Set of transmission lines which's from bus is $n$. |
| $L^{N,T(n)}$ | Set of candidate new lines which's to bus is $n$. |
| $L^{N,F(n)}$ | Set of candidate new lines which's from bus is $n$. |
| $P$ | Set of future time periods studied in TEP. |
| $R^{B(n)}$ | Set of renewables located on bus $n$. |
| $T$ | Set of time intervals in a day. |

*Parameters*

| | |
|---|---|
| $B^{\text{MVA}}$ | The MVA base of the grid model. |
| $C_g^G$ | The operation cost of generator g per MWh output. |
| $C_g^{G'}$ | The operation cost of new generator g per MWh output. |
| $C_g^{\text{On}}$ | The online cost of generator g. |
| $C_g^{\text{SU}}$ | The start-up cost of generator g. |
| $C_k^{L'}$ | The construction cost of line k. |
| $M$ | A large number M. |
| $N^P$ | Number of periods in the TEP study. |
| $N^Y$ | Number of years in each future period. |
| $N^T$ | Number of typical days in each year. |
| $N^D$ | Number of weekdays in a quarter. |
| $N^E$ | Number of weekend days in a quarter. |
| $N^B$ | Number of buses in the power system. |
| $N_{p,d,n}^H$ | Number of shedding hours on bus $n$ in typical day $d$ for future period $p$. |
| $p_g^{\text{Min}}$ | The minimum output power for generator $g$. |
| $p_g^{\text{Max}}$ | The maximum output power for generator $g$. |
| $p_{g,p}^{\text{Min}}$ | The minimum output power limit for new generator g in future period $p$. |
| $p_{g,p}^{\text{Max}}$ | The maximum output power limit for new generator g in future period $p$. |
| $p_r^{\text{Min,R}}$ | The minimum output power for renewable $r$. |
| $p_{r,t,d,p}^{\text{Max,R}}$ | The available power output for renewable $r$ at time interval $t$ in typical day $d$ for future period $p$. |
| $p_r^{\text{Min},R'}$ | The minimum output power for new renewable $r$. |
| $p_{r,t,d,p}^{\text{Max},R'}$ | The available power output for new renewable $r$ at time interval $t$ in typical day $d$ for future period $p$. |
| $p_{k,t}^{\text{Max}}$ | The active power rating of line $k$ at time interval $t$. |
| $p_{k,t,d,p}^{\text{Max}}$ | The active power rating of line $k$ at time interval $t$ in typical day $d$ for future period $p$. |
| $p_{k,t,d,p}^{\text{Max},L'}$ | The active power rating of candidate new line $k$ at time interval $t$ in typical day $d$ for future period $p$. |
| $p_{p,d,n,t}^{\square}$ | The load on bus $n$ at time interval $t$ in typical day $d$ for future period $p$. |
| $p_{p,d,n,t}^S$ | The load shedding of bus $n$ at time interval $t$ in typical day $d$ for future period $p$. |
| $p_{n,t,d,y}^D$ | The load demand on bus $n$ at time interval $t$ in typical day $d$ for future period $p$. |
| $p_{n,t}^D$ | The load demand on bus $n$ at time interval $t$. |
| $p_{r,t}^R$ | The output power of renewable power plant $r$ at time interval $t$. |
| $R_g^{10}$ | The reserve ramping rate of generator $g$. |
| $R_g^{\square}$ | The ramping rate of generator $g$. |
| $R^M$ | The ratio of the maintenance cost to the construction cost of a transmission line. |
| $x_k^{L'}$ | Reactance of candidate new line $k$. |

*Variables*

| | |
|---|---|
| $C^{OP}$ | The total operation cost of the grid in future periods. |
| $C^{CAP}$ | The total capital cost of the grid in future periods. |
| $p_{g,t,d,p}^G$ | The active power output of generator $g$ at time interval $t$ in typical day $d$ for future period $p$. |
| $p_{g,t,d,p}^{G'}$ | The active power output of new generator $g$ at time interval $t$ in typical day $d$ for future period $p$. |
| $p_{j,t,d,p}^L$ | The active power flow on line $j$ at time interval $t$ in typical day $d$ for future period $p$. |
| $p_{j,t,d,p}^{L'}$ | The active power flow on new line $j$ at time interval $t$ in typical day $d$ for future period $p$. |
| $p_{r,t,d,p}^R$ | The active power output of renewable $r$ at time interval $t$ in typical day $d$ for future period $p$. |
| $p_{n,t,d,p}^{R'}$ | The active power output of new renewables $r$ at time interval $t$ in typical day $d$ for future period $p$. |
| $p_{n,t}^S$ | The load shedding on bus $n$ at time interval $t$. |
| $p_{n,t}^{RC}$ | The renewable curtailment on bus $n$ at time interval $t$. |
| $p_{g,t}^{\square}$ | The output power of generator $g$ at time interval $t$. |
| $p_{k,t}^{\square}$ | The active power flow on line $k$ at time interval $t$. |
| $r_{g,t}^{\square}$ | The reserve of generator $g$ at time interval $t$. |
| $u_{k,p}^{L'}$ | When line $k$ is operational in future period $p$, its value is 1. Otherwise, its value is 0. |
| $v_{k,p}^{L'}$ | When line $k$ is constructed in future period $p$, its value is 1. Otherwise, its value is 0. |
| $\theta_{k,t}^{\square}$ | The phase angle difference of the two terminal buses of line $k$ at time interval $t$. |
| $\theta_{k,t,d,p}^F$ | The phase angle of the from bus of line $k$ at time interval $t$ in typical day $d$ for future period $p$. |
| $\theta_{k,t,d,p}^T$ | The phase angle of the to bus of line $k$ at time interval $t$ in typical day $d$ for future period $p$. |

## I. INTRODUCTION

The modern power system operations consider various economic and technical factors to achieve both cost-efficiency and physical system reliability [1]-[3]. Economic dispatch and unit commitment minimize the total generation cost while considering the power flows, generator production limits and other constraints [4]-[6]. They rely on the collection or prediction of various power system data such as demand, renewable production, and transmission network status [7]-[10]. Power systems need to expand to accommodate the rising demands, and the enhanced systems should possess the necessary transmission capability for reliable operation. New investments or upgrades in transmission and generation need to be well-planned to maintain the reliability of the power systems. The transmission expansion planning (TEP) typically looks several decades ahead due to the long construction time of transmission lines and the necessity to account for long-term shifts in grid load and generation. Generally, the TEP considers multiple scenarios of the future grids for the planning period, and the associated daily operation conditions under these scenarios are estimated and evaluated. Therefore, successful transmission planning relies on detailed and accurate predictions of the future grids at the nodal or facility level for load and generation. The loads are complicated and hard to precisely forecast in the long-term, due to various factors including weather and climate, socio-economics and electrification in transportation and other industry sectors [11]-[12]. The generation expansion including the locations and types of new generations is often a prerequisite for TEP. However, predicting the future generation investments at facility level is also complicated. Taking the U.S. as an example, different entities usually make their own decisions for developing new power plants [13]. After determining the generation expansion plan, the forecast of future renewable generation is also essential for a comprehensive TEP study on a renewable energy-dominated power grid [14].

The planning horizon for TEP usually spans decades, and it is widely acknowledged that climate changes will be more pronounced compared to long-standing historical patterns [15]-[16]. Climate change can affect multiple sectors of the future grids including load, generation, and transmission [17]-[19]. While the modern grids are becoming cleaner and greener, renewable production is highly dependent on the environmental variables such as wind speed, solar radiation, and temperature which may be affected by climate change. Besides, the electrical demands, especially the loads for heating, ventilation, and air conditioning, are highly correlated with the temperature [20]. Climate change may affect both the peak demand and the average demand. Moreover, the transmission line transfer capability is influenced by weather conditions including temperature and wind. While the dynamic line rating (DLR) technique is becoming more widely adopted in short-term grid operations, the impact of climate change on the transmission network should also be considered in long-term grid planning studies. Based on the above-mentioned reasons, the need to consider the weather and climate impact on future renewable energy-dominated grids in the TEP are raised.

The accurate and detailed future grid profiles are the foundation for TEP, and considering climate impact on various sectors of the grid may make the profile closer to the actual future conditions. Hence, data preparation is very important for TEP. However, few studies are presented for the methods and procedures to create the future grid profiles required for TEP. Most of the real-world power system data are sensitive and not publicly accessible. Instead, many synthetic test cases are available for research purposes, and some examples are IEEE/CIGRE benchmark cases [21]-[22], the PEGASE test case [23], and Polish Circle 2000 case [24]. Most of these cases represent the systems for a certain time snapshot only. While there are a handful of studies on the climate impact on the grids, very few studies present the spatio-temporal datasets of the future climate impacted grids which can be used by TEP. In this paper, we create the future profile of the Texas 123-bus backbone transmission (TX-123BT) power system [25], which includes the renewable production and line ratings influenced by meteorological variables. The representative profiles are also created and used as different scenarios in TEP. The datasets can be used to study the operation, planning, resilience, and many other analyses on climate-impacted grids. By utilizing this dataset, we are able to evaluate the performance improvement of TEP after considering the climate impact.

The trend of decarbonization in the energy field may lead to the change in both generation and consumption sides. Due to incentive policies and other factors, more variable renewable energy (VRE) will be invested and deployed into the future power system. Meanwhile, conventional generation like coal and gas plants are expected to decline. According to the National Renewable Energy Laboratory, 70% of U.S. total energy is expected to be generated by renewables by 2035, and it will rise to 90% by 2050 [26]. Expected high renewable penetration requires improving current TEP strategies to handle VRE's versatility and uncertainties. Therefore, the U.S. Department of Energy (DOE) are proposing the state-of-the-art TEP techniques to address challenges arising from high renewable penetration [27]. Besides, climate change will directly influence the weather conditions in future, and thus influence renewable production. The impact may also increase the versatility and uncertainties of VRE, and thus increase the difficulties of operation and planning of future grids. An intuitive method to adapt to the versatility is increasing the resolution of the VRE in the TEP model. A TEP model considering climate impact (TEP-CI) with higher resolution for renewable production and line flow limits are proposed in this paper. With TEP-CI, we can have an initial study on the planning of renewable climate-impacted grids. It can also serve as a benchmark for developing more comprehensive TEP-CI models.

To numerically evaluate the performance of the TEP-CI, we developed a security-constrained unit commitment (SCUC) model that can reflect the future grid investment and climate change impacts and be used to simulate the daily operations of the future grids. The SCUC simulations are conducted on all typical days to obtain the operation conditions of the future grid in different planning epochs. Three widely-used reliability indices: loss of load probability (LOLP), loss of load expectation (LOLE), and expected unserved energy (EUE) are used to evaluate the overall reliability of the grid for each certain future period. We showcase how different types of investment will influence the future grid by conducting the reliability analysis in the following three situations respectively: (i) a grid without any asset investment, referred as future (FR) case; (ii) a grid with only future generation investment (FGI); and (iii) a grid with both future generation and transmission investment (FGTI). The future power system reliability under different investment situations is analyzed.

In the literature, very few studies address the creation of accurate future grid profiles, or perform numerical analysis on the impact of weather and climate on the renewable energy-dominated grids and how it will influence the transmission planning. In [28], the climate impact on various components of power system is discussed and concluded, however no numerical analysis is conducted. In [29], the future planning of the climate-impacted Indonesia power system is studied. However, the system consists of less than 20% renewable resources, and the effects of climate impacts on transmission network are not considered. In [30], the climate impact on generation mix of Portuguese power systems is studied. Since the generation is not studied at facility level, the study cannot capture the spatio-temporal characteristics of the climate impacts on energy production.

Many studies focus on the modeling of transmission expansion considering the specific impact of the weather or climate conditions. In [31], the stochastic TEP model includes the dynamic and uncertain nature of line ratings. The economic benefits obtained by considering the DLR are validated, but the reliability of the future grids with its obtained investments are not investigated. A three-stage robust model that considers the climate impacts on renewable production is proposed in [32]. The model accounts for the effects of El Niño and La Niña on renewable production but may not fully capture the nuances of climate impact on a regional scale. The model in [33] addresses the uncertainties of renewable production. However, it does not consider climate impact on renewable production and transmission line capacity.

Based on the above literature, current transmission planning methodologies utilize the predicted information of future representative scenarios which do not consider the climate impact. Existing models do not consider higher temporal resolution and capture temporal changes for these grid conditions. Existing evaluation of transmission planning focuses on economic aspects. The reliability of the future grid, especially after suitable transmission investment, remains to be investigated. Few studies conclude and present the whole procedures for transmission planning, including data preparation, model formulation and numerical evaluation. To the best of the authors' knowledge, this work is the very first to address the climate impact on all these techniques and procedures. Various meteorological variables are considered in future renewable energy generation and line rating profiles. The proposed TEP-CI model considers the fluctuation of weather-dependent renewable production and dynamic line rating. A specialized SCUC model which incorporates climate-dependent variations and load shedding is also required to obtain the future grid daily operation conditions for the reliability evaluations. The reliability indices which can evaluate grid reliability in a long future period are calculated and compared. We evaluate the reliability of the grid planned with the proposed TEP-CI model to study the necessity and performance improvement of considering climate change in transmission planning. Below are the main contributions of this paper,

- The climate-impacted profiles of the TX-123BT from 2020 to 2050 including renewable power production and dynamic line ratings are created. The representative profiles are also created for the planning and other scenario-based studies.
- The TEP-CI considers the versatility of renewable and climate impact, and is improved to adapt the spatio-temporal data of the representative profiles.
- Three reliability indices (LOLP, LOLE, EUE) are introduced to evaluate the long-term grid reliability under various investment situations and TEP models.

The rest of this work is structured as follows. Section II presents the procedures to create the time-sequential climate-impacted future power system profiles. Section III shows the improved TEP model for climate-impacted renewable energy-dominated grids. The SCUC model and simulations are presented in section IV, while the reliability evaluation methods and results are shown in Section V. The conclusions are drawn in Section VI.

II. CLIMATE IMPACTED GRID PROFILES

Typically, the transmission planning needs to consider the future power system operation conditions under different sce-narios in future periods. Thus, the future gird profiles including forecast load and generation information are critical for TEP to give a suitable transmission investment plan. The transmission planning requires both comprehensive technical data of the current grid configurations and future prospective information. Our methodology requires comprehensive geographic details of the grid infrastructure and dependable forecasts of future climate conditions specific to the grid's region to create future grid profiles that incorporate climate impacts.

The TX-123BT is a synthetic power system based on the footprint of Texas [34]. It is designed to represent the Electric Reliability Council of Texas (ERCOT) system, which covers most areas in Texas territory. This system includes 345 KV high-voltage transmission network distributed in ERCOT, which is shown in Fig. 1. The TX-123BT system's generator capacities and load distributions are closely aligned with the actual ERCOT as it was in 2019. Compared with other publicly accessible test power systems, TX-123BT provides the geographic locations of all substations, transmission lines, thermal generators, and renewable power plants. These geographic locations are necessary to obtain the future climate and weather conditions at the facilities, which are required for the creation of renewable production and dynamic line rating profiles. We create the climate-impacted power system profiles based on the future climate data extracted from Coupled Model Intercomparison Project Phase 6 (CIMP6) for 2020-2050. The CIMP6 is an advanced, comprehensive, coupled model global climate change project [35]. CIMP6 generates climate projections based on a variety of scenarios using complex climate models. The projections are based on a range of scenarios called Shared Socioeconomic Pathways (SSPs) combined with Representative Concentration Pathways (RCPs). SSPs describe possible future changes in demographics, economics, technology, energy consumption, and land use. RCPs outline pathways of greenhouse gas concentrations and their radiative forcing on the climate system. By combining SSPs with RCPs, CMIP6 explores a wide range of future climate outcomes. The extracted data are for meteorological variables such as wind speed, solar radiation, and temperature under Representative Concentration Pathway (RCP) 8.5, which is considered as the most likely global warm conditions if the world makes usual efforts on emission reduction in future [36]. We compare the forecast climate data from CIMP6 model with the historical climate data from North American Land Data Assimilation System (NLDAS-2) [37] for the same period 2019-2022, and we verify that these meteorological variables in CIMP6 is coherent with the historical observations. Fig. 2. displays a year-long comparison of temperature data from the two datasets at a bus location.

The CIMP6 climate data has 3-hour resolution. In each 3-hour period, the wind speed, solar radiation, and temperature data for all bus locations in TX-123BT are extracted. Based on the weather-dependent models for dynamic line rating, solar production and wind production in [34], the corresponding profiles are created for 2019-2050, and have the same 3-hour resolution. Our renewable production and line ratings profiles are calculated for each renewable power plant or transmission line at three-hour resolution, based on the future climate forecast at the specific location of the facility.

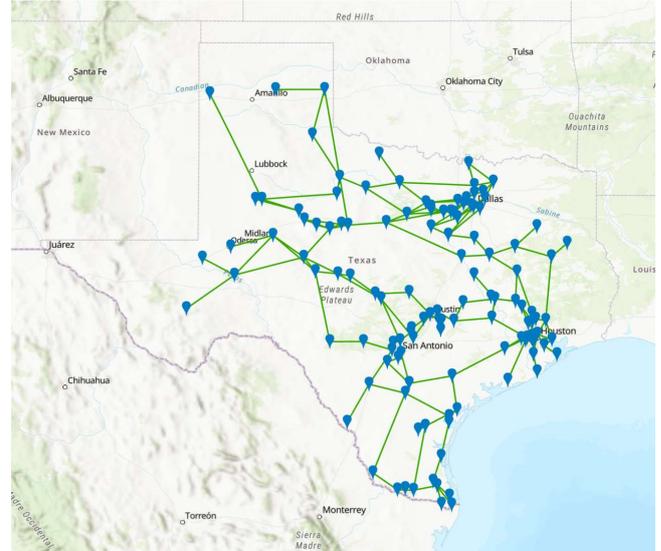

Fig. 1. TX-123BT Backbone Transmission Network [34].

For dynamic line rating, the lower wind speed, higher temperature and solar radiation on the two terminal buses of the transmission line are averaged respectively and then used in the calculation. The monthly average line ratings of a transmission line for 2019-2024 are plotted and shown in Fig. 3.

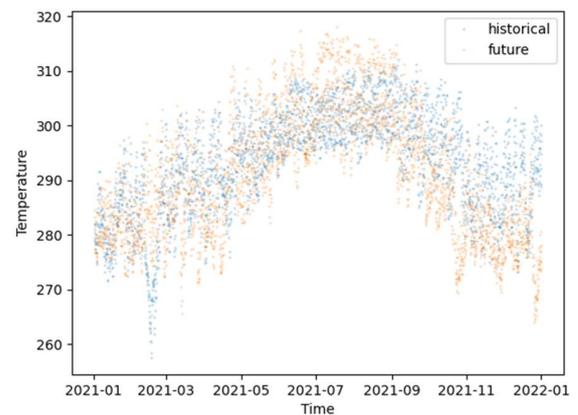

Fig. 2. Comparison of Temperature Data in CIMP6 and NLDAS-2.

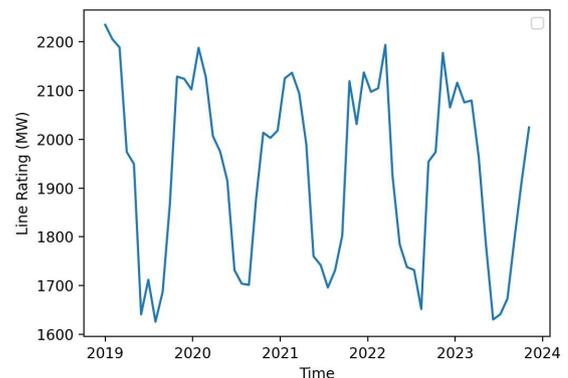

Fig. 3. Plot of the Averaged Dynamic Line Rating at Line 1.

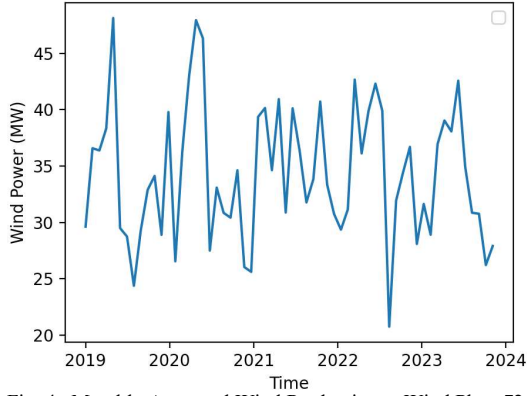

Fig. 4. Monthly Averaged Wind Production at Wind Plant 72.

The gross power output for a wind farm is the aggregation of all the wind turbines in it. To simplify the relationship between the wind speed at the wind farm location and wind farm power output, we assume the wind turbines in one wind farm are the same type. Besides, the wind speed at the wind turbine height is required for wind power production calculation. Since the wind speed in CIMP6 is the wind speed at the 'surface' of earth at 10m height, we estimate the wind speed at 80m using the log wind profile [38]. The logarithmic wind profile model is based on the assumption that wind speed increases with altitude due to the decrease in surface drag. This method provides a reliable approximation of wind speed at the turbine hub height, which is essential for our wind production profile creation based on the CIMP6 dataset. The logarithmic relationship between wind speeds at 80m height and 10m above the surface is:

$$v_{80} = v_{10} \times \frac{\ln\left(\frac{80}{z_0}\right)}{\ln\left(\frac{10}{z_0}\right)} \quad (1)$$

where $z_0$ represents the surface roughness length, typically ranging from 0.03 m for smooth, open surfaces to 1.0 m for rough, vegetated terrain. Based on the estimated wind speed and the wind power production model, the wind profiles are created. The monthly average wind power production of a wind plant is plotted in Fig. 4.

The solar production is calculated using both shortwave and longwave radiation data extracted from CIMP6. The effective radiation on the solar panel is estimated based on the frequency range of the commonly used solar panels. The solar production of solar plant 66 is plotted and shown in Fig. 5.

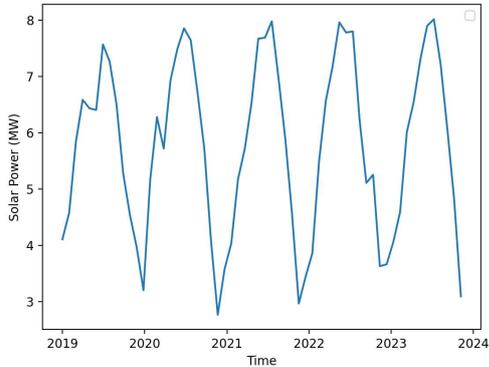

Fig. 5. Monthly Averaged Solar Production at Solar Plant 66.

Due to the computation burden, future power system planning models such as TEP often consider a limited number of representative future scenarios. Hence, we create the representative daily profiles for every quarter of each 5-year period from 2021 to 2050. Each representative profile encapsulates the average renewable production and line ratings derived from identical hours across all days within the same quarter. Since electric demand is significantly affected by social activities, we establish representative load profiles for weekdays and weekends within each quarter. The line rating, wind production, solar production, and load profiles for quarter 1, 2021-2025 are shown in Fig. 6(a)-(d).

In Fig. 6(a), the line rating in Quarter I is higher than other quarters due to the low temperature. In Quarter III, the line rating is the lowest and drops obviously during noon time due to the high temperature. According to Fig. 6(b)-(c), the wind production is obviously lower in summer, while load is obviously higher. The plots illustrate why summer and winter (high wind and low load) scenarios are necessary to be both analyzed in some industrial applications in Electric Reliability Council of Texas (ERCOT).

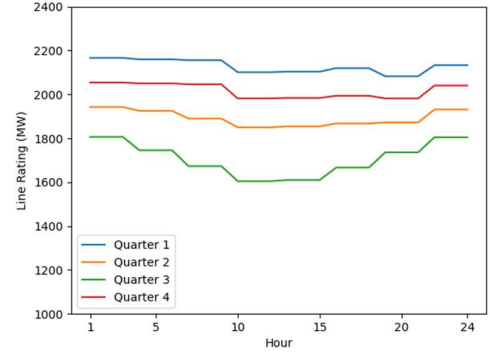

Fig. 6(a). Representative Line Ratings of a transmission Line in 2021-2025.

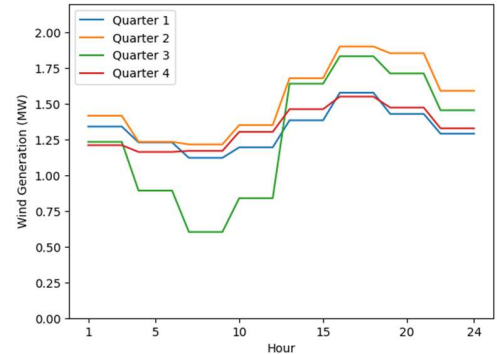

Fig. 6(b). Representative Power Production of a Wind Plant in 2021-2025.

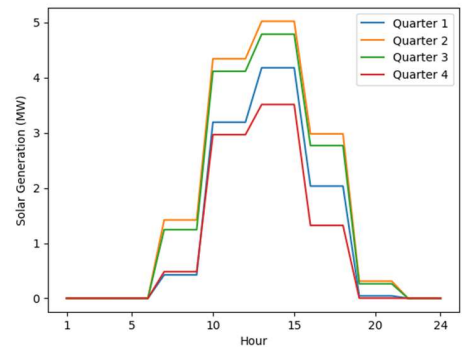

Fig. 6(c). Representative Power Production of a Solar Plant in 2021-2025.

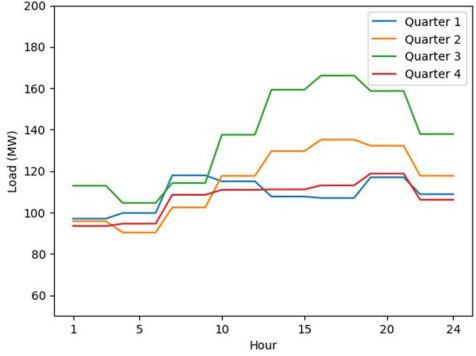

Fig. 6(d). Representative Loads on a Bus on weekday, 2021-2025.

Since the created representative profiles don't include the future new invested power plants and transmission lines, they will be regarded as the benchmark case for the TEP simulation, referred to as TX-123BT Future (FR) case. The new power plants, including renewable resources, are interconnected with the power system through queue systems by various entities in United States. An agent-based model (ABM) is used to mimic the generation investment behavior by the market participators [39]. Each market participator is regarded as an agent and can make their own investment decisions based on the market and grid operation information. The ABM was selected due to its unique ability to model the complex interactions and decision-making processes among various stakeholders in the energy market, including utility companies, independent power producers, and regulatory bodies. Unlike traditional modeling approaches that might simplify these interactions through aggregated supply-demand curves or static investment decisions, ABM allows for a dynamic representation of how individual decisions and actions can lead to emergent market behaviors and investment patterns. In [39], the generation investment obtained from ABM is analyzed and validated. The future generation investment for TX-123BT including the capacity and plant type of different market participator, are obtained based on the ABM model. Then, the future production of the new renewables is calculated. The future TX-123BT with new generation investment and related profiles is named as the TX-123BT future generation investment (FGI) case.

### III. TEP-CI FOR CLIMATE IMPACTED GRIDS

Traditionally, the TEP model considers the future grid and market trends such as increasing loads and fuel price, whereas the weather and climate impact is not considered. The climate may impact the grid on various sectors such as transmission, generation, and loads. In section II, the representative profiles including the renewable production and dynamic line ratings of the future climate impacted grid are created. In this paper, we address the timely changing characteristics of the renewable production and dynamic line ratings due to the meteorological variables in the TEP model. In this section, the modeling of the proposed TEP-CI and its transmission investment plan on the TX-123BT are presented.

#### A. Modeling of TEP-CI

In order to consider these climate-impacted components in the TEP-CI, we have made the following updates to the TEP model: i) The line flow capacity constraints now include the changing line ratings for different hours in the representative scenario; ii) The constraints are added to describe the available renewable resources in the system; iii) The power balance equation is modified to include the weather-dependent renewable production and load profiles. After the update, the TEP-CI can utilize the representative profiles created in Section II. The detailed TEP-CI model is shown below.

$$\min C^{OP} + C^{CAP} \tag{2}$$

$$C^{OP} = N^Y \cdot B^{MVA} \cdot \frac{365}{N^D} \cdot \sum_{g \in G, t \in T, d \in D^T, p \in P} p^G_{g,t,d,p} \cdot C^G_g$$

$$+ N^Y \cdot B^{MVA} \cdot \frac{365}{N^T} \cdot \sum_{g \in G', t \in T, d \in D^T, p \in P} p^{G'}_{g,t,d,p} \cdot C^{G'}_g \tag{3}$$

$$C^{CAP} = \sum_{k \in L', p \in P} V^{L'}_{k,p} \cdot C^{L'}_k (1 + (N^P - p + 1)R^M \cdot N^Y) \tag{4}$$

$$\sum_{j \in L^{T(n)}} p^L_{j,t,d,p} - \sum_{j \in L^{F(n)}} p^L_{j,t,d,p} + \sum_{j \in L^{N,T(n)}} p^{L'}_{j,t,d,p}$$
$$- \sum_{j \in L^{N,F(n)}} p^{L'}_{j,t,d,p} + \sum_{g \in G^{B(n)}} p^G_{g,t,d,p} + \sum_{g \in G^{N,B(n)}} p^{G'}_{g,t,d,p}$$
$$+ \sum_{r \in R^{B(n)}} p^R_{r,t,d,p} + p^{R'}_{n,t,d,p} = p^D_{n,t,d,p} \tag{5}$$
$$\forall n \in B, t \in T, d \in D^T, p \in P$$

$$p^{Min}_g \leq p^G_{g,t,d,p} \leq p^{Max}_g \tag{6}$$
$$\forall g \in G, t \in T, d \in D^T, p \in P$$

$$p^{Min}_{g,p} \leq p^{G'}_{g,t,d,y} \leq p^{Max}_{g,p} \tag{7}$$
$$\forall g \in G', t \in T, d \in D^T, p \in P$$

$$p^{Min,R}_r \leq p^R_{r,t,d,p} \leq p^{Max,R}_{r,t,d,p} \tag{8}$$
$$\forall r \in R, t \in T, d \in D^T, p \in P$$

$$p^{Min,R'}_r \leq p^{R'}_{r,t,d,y} \leq p^{Max,R'}_{r,t,d,p} \tag{9}$$
$$\forall r \in R', t \in T, d \in D^T, p \in P$$

$$-p^{Max}_{k,t,d,p} \leq p^L_{k,t,d,p} \leq p^{Max}_{k,t,d,p} \tag{10}$$
$$\forall k \in L, t \in T, d \in D^T, p \in P$$

$$-M(1 - u^{L'}_{k,p}) \leq p^{L'}_{k,t,d,p} - \frac{\theta^F_{k,t,d,p} - \theta^T_{k,t,d,p}}{x^{L'}_k} \leq$$
$$M(1 - u^{L'}_{k,p}), \forall k \in L', t \in T, d \in D^T, p \in P \tag{11}$$

$$-p^{Max,L'}_{k,t,d,p} \cdot u^{L'}_{k,p} \leq p^{L'}_{k,t,d,p} \leq p^{Max,L'}_{k,t,d,p} \cdot u^{L'}_{k,p}) \tag{12}$$
$$\forall k \in L', t \in T, d \in D^T, p \in P$$

$$\sum_{p' \in P, p' \leq p} u^{L'}_{k,p'} \leq u^{L'}_{k,p} \quad \forall k \in L', p \in P \tag{13}$$

$$v^{L'}_{k,p} \geq u^{L'}_{k,p} - u^{L'}_{k,p-1}, \forall k \in L', p \in P, p > 1 \tag{14}$$

$$v^{L'}_{k,1} = u^{L'}_{k,1} \quad \forall k \in L' \tag{15}$$

The TEP-CI can minimize the operation and transmission investment cost for the studied period by (2). The operation cost includes both the existing and new thermal generators in (3). The capital cost of the transmission line is simplified by assuming the yearly maintenance cost is a partial of the total construction cost in (4). The nodal power balance addresses the available renewable resources for each time interval by (5). The power output constraint for existing thermal generators and newly invested generators obtained from the ABM are described by (6) and (7) separately. The renewable power output for each time interval should be under its available amount. To be noticed, $p_{r,t,d,p}^{Max,R}$ is the maximum available renewable output in the representative profiles, which is calculated using the CIMP6 climate data. The power output constraints for both existing and new renewables are (8)-(9). The line flow limit for existing transmission line is in (10), and $p_{k,t,d,p}^{Max}$ is the dynamic line ratings in the representative profiles. To model the line flow limit of the new transmission line, we use a big number M and the binary variable $u_{k,p}^{NL}$ to enforce the DC power flow constraint when the line is constructed, as described in (11). The flow limit of the new lines is described by (12). The constraints related to the binary variables for line construction are shown in (13)-(15).

While this model integrates a wide range of general physical constraints related to power flow and generators, it demands accurate forecasts for future renewable energy production and dynamic line ratings affected by weather variations. The reliability of these forecasts hinges on specific climate models that may not encompass every potential future climate scenario, along with detailed geographic data at the facility level for the transmission network and renewable power plants.

*B. Transmission Investment Plan from TEP-CI*

The TEP models are implemented using Python with Pyomo package [41]. Since the models are formulated as mixed-integer linear programming (MILP) problem, the commercial solver Gurobi [42] is used to find optimal solutions for the models. Our proposed TEP model is designed to accommodate various resolution profiles, such as one-hour or three-hour intervals, denoted by *t* periods within a day. Given that our profiles are created at a 3-hour resolution, preserving this granularity in the model helps to shorten simulation duration without compromising the quality of the solutions. In the first TEP-CI simulation, we set the number of year-epoch to 3, and each epoch represents a 5-year period. Thus, it determines the transmission planning of TX-123BT for the period 2021-2035. Even for the TX-123BT FR case which has no generation investments made during this planning period, the TEP-CI can easily find a feasible solution. The transmission line investment results are shown in Table II. The total cost, generation costs and transmission line investment cost are presented in Table I.

The TEP-CI finds 15 transmission lines to be invested. The transmission line investment cost is $3B, which is about 2.94% of the total cost for TX-123BT FR during 2021-2035.

In the second TEP-CI simulation, we set the number of year-epoch to 6, which means the TEP-CI will give the transmission investment results for 2021-2050. For this future period, the safety-secured operation cannot be maintained without load shedding for FR case. The electricity demands are expected to increase rapidly in 2035-2050, and existing generation resources cannot meet the needs for such large amounts of loads. Using the representative profiles under FGI case as input for TEP-CI, the simulation results are shown in Table III and IV.

TABLE I
TEP INVESTMENT AND SYSTEM OPERATION COSTS IN 2021-2035 FOR FR CASE

| Costs | Amount ($) |
|---|---|
| Generation Cost | 102.87B |
| Transmission Line Investment Cost | 3.02B |
| Total Cost | 105.89B |

TABLE II
NEW TRANSMISSION LINES INVESTMENT IN 2021-2035 FOR FR CASE

| New Line Number | Construction Period | New Line Number | Construction Period |
|---|---|---|---|
| 2 | 2021-2025 | 68 | 2025-2030 |
| 6 | 2026-2030 | 72 | 2021-2025 |
| 7 | 2021-2025 | 74 | 2026-2030 |
| 22 | 2025-2030 | 165 | 2026-2030 |
| 31 | 2021-2025 | 171 | 2026-2030 |
| 50 | 2025-2030 | 233 | 2031-2035 |
| 56 | 2025-2030 | 249 | 2021-2025 |
| 58 | 2025-2030 | | |

TABLE III
TRANSMISSION LINES INVESTMENT IN 2021-2050 FOR FGI CASE

| New Line Number | Construction Period | New Line Number | Construction Period |
|---|---|---|---|
| 3 | 2036-2040 | 80 | 2041-2045 |
| 6 | 2041-2045 | 74 | 2021-2025 |
| 7 | 2031-2035 | 82 | 2041-2045 |
| 8 | 2036-2040 | 83 | 2021-2025 |
| 9 | 2031-2035 | 112 | 2036-2040 |
| 30 | 2041-2045 | 147 | 2021-2025 |
| 49 | 2041-2045 | 189 | 2021-2025 |
| 57 | 2041-2045 | 191 | 2021-2025 |
| 72 | 2021-2025 | 247 | 2021-2025 |

TABLE IV
THE TEP INVESTMENT AND SYSTEM OPERATION COSTS (2021-2050)

| Costs | Amount ($) |
|---|---|
| Generation Cost | 146.76B |
| Transmission Line Investment Cost | 6B |
| Total Cost | 152.76B |

We calculate the dynamic ratings of the new lines in the investment plan, for the future periods after they are constructed. The new transmission lines and their dynamic rating profiles are then integrated into the TX-123BT FGI case and form the future scenario that includes both generation and transmission investments. It is called the TX-123BT future generation and transmission investment (FGTI) case.

The comparisons between TEP-CI and traditional TEP method [40] are concluded in Table V. TEP-CI results in greater transmission investments due to its utilization of detailed and high-resolution climate-impacted profiles. While this approach leads to higher initial transmission costs, it significantly lowers generation costs over the planning periods.

TABLE V
TEP AND TEP-CI RESULTS FOR FGI CASE IN 2021-2035

|  | TEP | TEP-CI |
|---|---|---|
| Total Costs | $ 79.34B | $ 55.73B (-29.7%) |
| Transmission Line Investment Costs | $ 0.24B | $ 0.47B (+ 95.8%) |
| Generation Costs | $ 79.1B | $ 55.26B (-30.1%) |
| Total Flow on Transmission Lines | 18,727 MWh | 22,988 MWh (+22.75) |

Table VI presents the solving times for TEP-CI compared to traditional TEP models. For shorter planning spans, such as 15 years, TEP-CI does not significantly extend solving times. However, due to TEP-CI model considers higher resolution of climate-impacted profiles, TEP-CI solving times increase markedly for longer spans.

TABLE VI
THE SOLVING TIME (SECONDS) FOR TEP-CI AND TRADITIONAL TEP

| Planning Span | TEP | TEP-CI |
|---|---|---|
| 15 Years | 209.5 | 214.7 (+2.48%) |
| 30 Years | 6235.6 | 8821.1 (+41.4%) |

In this paper, we present TEP-CI simulations and results specifically for the TX-123BT model. Adapting our methodology to power systems across various geographic regions necessitates precise predictions for future renewable energy production and dynamic line rating profiles. Achieving such accuracy may involve integrating region-specific climate and weather forecast data and utilizing the ABM approach or other regionally appropriate methods, to accurately forecast future renewable energy investments.

## IV. SCUC FOR TEP EVALUATION

The SCUC model is modified and customized specifically for the future climate-impacted study and TEP performance evaluation. Firstly, to study the reliability performance of the grid under different investment plans, also due to the potential reliability issues in the future grid, the load shedding should be considered in the SCUC model. Specifically, the load shedding variables are introduced to the power balance equations, and the constraints describing the maximum shedding amount are added to the SCUC model. The loads will be shedded only when the power systems cannot be operated safely. The shedding should happen when the physical constraints cannot be satisfied, because it may cause both economic losses and social disturbance. Hence, a penalty term is added to the objective function of the SCUC model, to make sure the load shedding can only happen when it is necessary. With the improved model, the SCUC solutions can give us information about the unserved load amount due to the electrical demand increasing and climate change in the future.

Secondly, the climate-dependent grid profiles have the same 3-hour resolution as the climate data, while the commonly used SCUC has hourly resolution. We can simply transfer the 3-hour resolution profiles into hourly profiles by assuming all the 3 hours have the same data and used as hourly data input for SCUC. However, for both SCUC simulations on a large number of future profiles and the TEP for climate-impacted grids, the number of time intervals in models will significantly influence the computation time. Hence, the SCUC model is adjusted from hourly resolution to 3-hour resolution. Besides, the SCUC input data such as generator costs c0, c1 and generator ramping rate data are updated for 3-hour resolution. The detailed formulation of SCUC for TEP evaluation is shown below.

$$\min \sum_{g \in G(p)} \sum_{t \in T} (c_g P_{g,t} + c_g^{On} u_{g,t} + c_g^{SU} v_{g,t})$$
$$+ M \cdot \sum_{b \in B} \sum_{t \in T} p_{n,t}^S \quad (16)$$

$$p_{n,t}^{RC} = 0, \quad \forall (n,t) \in ST^1 \quad (17)$$

$$ST^1 = \{(n,t) \mid n \in B, t \in T, s.t. \ p_{n,t}^D - \sum_{r \in R(b)} p_{r,t}^R \geq 0\} \quad (18)$$

$$p_{n,t}^{RC} \leq \sum_{r \in R(b)} p_{r,t}^R - p_{n,t}^D, \quad \forall (n,t) \in ST^2 \quad (19)$$

$$ST^2 = \{(n,t) \mid n \in B, t \in T, s.t. \ p_{n,t}^D - \sum_{r \in R(n)} p_{r,t}^R < 0\} \quad (20)$$

$$p_{n,t}^{RC} \leq p_{n,t}^D - \sum_{r \in R(n)} p_{r,t}^R \quad \forall (n,t) \in ST^1 \quad (21)$$

$$p_{n,t}^{RC} = 0 \quad \forall (n,t) \in ST^2 \quad (22)$$

$$P_g^{min} \cdot u_{g,t} \leq P_{g,t} \quad \forall g,t \quad (23)$$

$$P_{g,t} + r_{g,t} \leq P_g^{max} u_{g,t} \quad \forall g,t \quad (24)$$

$$0 \leq r_{g,t} \leq R_g^{10} u_{g,t} \quad \forall g,t \quad (25)$$

$$\sum_{m \in G} r_{m,t} \geq P_{g,t} + r_{g,t} \quad \forall g,t \quad (26)$$

$$-R_g \leq P_{g,t} - P_{g,t-1} \leq R_g \quad \forall g,t \quad (27)$$

$$P_{k,t} = \theta_{k,t}/x_k \quad \forall k,t \quad (28)$$

$$-P_{k,t}^{max} \leq P_{k,t} \leq P_{k,t}^{max} \quad \forall k,t \quad (29)$$

$$\sum_{g \in G(n)} P_{g,t} + \sum_{k \in K(n-)} P_{k,t} - \sum_{k \in K(n+)} P_{k,t} + \sum_{r \in R(n)} p_{r,t}^R$$
$$= p_{n,t}^D + p_{n,t}^{RC} - p_{n,t}^S \quad \forall n,t \quad (30)$$

$$v_{gt} \geq u_{gt} - u_{g,t-1} \quad \forall g, t > 1 \quad (31)$$

In (16), the SCUC will optimize the operation cost for the day, and an additional term is added to ensure load shedding is employed strictly as a last resort. There will be no renewable curtailment on a bus when the total renewable power is less than the load at the location according to (17)-(18). The maximum renewable curtailment is constrained by (19)-(20). Similarly, the load shedding can only be made when the renewable power on the bus is not sufficient in (21)-(22). The thermal generator minimum and maximum power output, the reserve constraints, and ramping limits are in (23)-(27). The DC power flow and line flow limits are in (28)-(29). The nodal power balance equation includes both the load shedding and renewable curtailments by (30). The generator online and starting binary variables are constrained by (31).

In each FR, FGI and FGTI case, the future representative profiles from 2021 to 2050 includes 48 daily profiles, for weekdays and weekends in each quarter in each 5-year planning epoch. The SCUC simulations are run on all the daily profiles for different scenarios. The weekday highest load shedding for different quarters in 2041-2045, 2046-2050 are shown in Fig. 7.

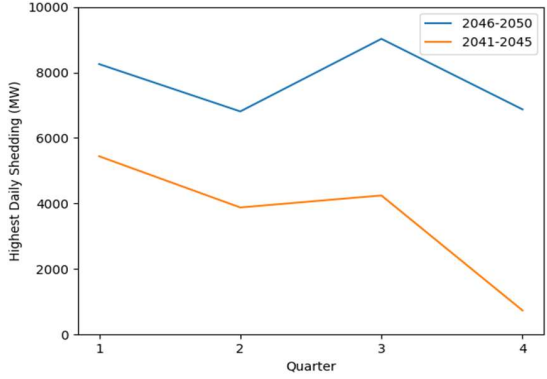

Fig. 7. Highest Daily Load Shedding for Weekdays Under FR Scenario.

We observe that under FR scenario, the grid must necessarily shed loads after 2040. The required load shedding amount grows rapidly due to the increasing load demand. It indicates that the current system conditions cannot handle the increasing loads in 20 years later. It is reasonable given the absence of expansions and developments in generation and transmission infrastructure under FR scenario.

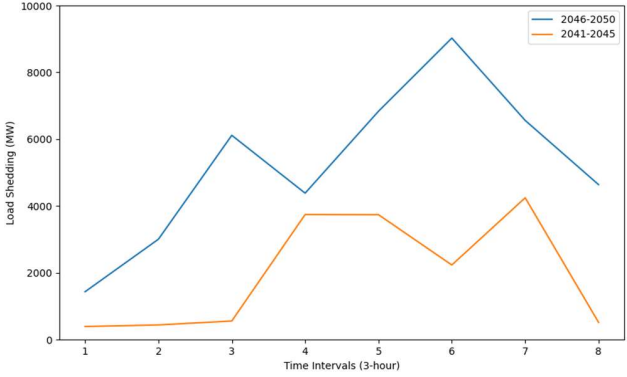

Fig. 8. Load Sheddings for Quarter 3 weekends in 2041-2045 and 2046-2050.

The load sheddings for weekends in Quarter 3 in periods 2041-2045 and period 2046-2050 under FR scenario are plotted in Fig. 8. Load shedding for 2046-2050 is substantially greater than that of 2041-2045, and the daily shedding patterns differ significantly between the two five-year spans.

TABLE VII
TOTAL OPERATION COSTS AND GENERATION IN QUARTER 3 UNDER FGTI SCENARIO FOR DIFFERENT 5-YEAR PERIODS

| 5-year Period | Total Operation Costs (M $) | | Total Generation (GWh) | |
| --- | --- | --- | --- | --- |
| | Weekdays | Weekends | Weekdays | Weekends |
| 2021-2025 | 11.55 | 11.62 | 711.1 | 714.2 |
| 2026-2030 | 13.94 | 13.90 | 810.27 | 808.6 |
| 2031-2035 | 17.41 | 15.31 | 940.3 | 858.5 |
| 2036-2040 | 21.52 | 17.90 | 1090.7 | 956.4 |
| 2041-2045 | 26.83 | 21.18 | 1263.9 | 1070.2 |
| 2046-2050 | 33.43 | 25.17 | 1462.1 | 1199.1 |

Under the FGTI scenario, Table VII summarizes the daily operation costs and total scheduled generation from thermal plants. A comparison between the weekdays in periods 2021-2025 and 2046-2050 reveals that the total generation approximately doubled, while the total operation costs have roughly tripled. This discrepancy is attributable to the higher marginal prices associated with generator's high power output level. This situation arises because few thermal power plants are invested in the future, while the total generation is increasing.

## V. RELIABILITY EVALUATION

To evaluate the reliability of future grids under different investment scenarios, we develop several reliability indices which can be calculated based on SCUC numerical results of the operation conditions for all days in future. We select the concepts of reliability indices which are widely used in industry and academics as the indices for our risk analysis. These indices include the expected unserved energy (EUE), loss of load probability (LOLP) and loss of load expectation (LOLE). The three indices are the implications of the power system's ability to reliably meet energy demand from different aspects. Each index offers a unique lens through which the robustness and resilience of the grid can be assessed, catering to the intricate dynamics of power supply and demand. Expected unserved energy (EUE) quantifies the total energy that cannot be supplied due to system limitations within a specified timeframe, serving as a direct measure of the magnitude of energy deficit. A higher EUE indicates more significant instances where the grid fails to meet demand, pointing towards potential weaknesses in generation capacity or transmission infrastructure. This index is especially crucial in evaluating the system's performance during peak demand periods or in scenarios with high renewable energy variability, where the balance between supply and demand is most delicate. Loss of load probability (LOLP) assesses the likelihood that the power system will not meet the demand at any given time, essentially reflecting the system's overall reliability. An increase in LOLP suggests a greater risk of power shortages, signaling the need for enhanced system planning and investment in reliability improvements. It underscores the importance of having sufficient reserve margins and flexible resources that can quickly respond to fluctuations in demand and supply. Loss of load expectation (LOLE), expressed in hours per year, estimates the expected duration of load not being served. This index complements LOLP by providing insight into the length of time the system might be under stress, thus affecting consumer experience and economic activities. A lower LOLE value is indicative of a power system that, while it may occasionally fail to meet demand, does so for a minimal duration, minimizing disruption to end-users.

In synthesizing the insights derived from EUE, LOLP, and LOLE, decision-makers and stakeholders gain a comprehensive understanding of the power system's operational challenges and areas requiring attention.

The EUE is an index that can evaluate the amount of total unserved energy for a given period, such as one year. It can evaluate the scale of the outage by calculating the total unserved energy amount.

$$\text{EUE} = \sum_{p \in P, d \in D^{T(D)}, n \in B, t \in T} p^{\text{S}}_{p,d,n,t} \cdot N^{\text{D}} \\ + \sum_{p \in P, d \in D^{T(E)}, n \in B, t \in T} p^{\text{S}}_{p,d,n,t} \cdot N^{\text{E}} \quad (32)$$

The LOLP is the probability of load loss / shedding occurrence. Specifically, it measures how often the power system cannot serve all the loads, such as load curtailment or black out. The LOLP is usually calculated for a specific period, such as one year.

$$\text{LOLP} = \text{EUE}/(\sum_{p \in P, d \in D^{T(D)}, n \in B, t \in T} p_{p,d,n,t}^{\square} \cdot N^{\text{D}} + \sum_{p \in P, d \in D^{T(E)}, n \in B, t \in T} p_{p,d,n,t}^{\square} \cdot N^{\text{E}}) \quad (33)$$

TABLE VIII
DIFFERENT RISK INDICES OF FUTURE TX-123BT WITHOUT INVESTMENT

| Index | P1 | P2 | P3 | P4 | P5 | P6 |
|---|---|---|---|---|---|---|
| Annual LOLP | 0% | 0% | 0% | 0.013% | 0.81% | 4.06% |
| LOLE (Hours/Bus) | 0 | 0 | 0 | 2.92 (0.036%) | 25.36 (0.32%) | 125.26 (1.55%) |
| EUE (MWh) | 0 | 0 | 0 | 69,105.24 | 4,803,316 | 27,233,362 |

P1 to P6 denotes the future periods from 2021-2025 to 2046-2050.

TABLE IX
DIFFERENT RISK INDICES OF FUTURE TX-123BT WITH TEP-CI

| Index | P1 | P2 | P3 | P4 | P5 | P6 |
|---|---|---|---|---|---|---|
| Annual LOLP | 0% | 0% | 0% | 0% | 0.0716% | 2.56% |
| LOLE (Hours/Bus) | 0 | 0 | 0 | 0 | 2.92 (0.036%) | 100.48 (1.24%) |
| EUE (MWh) | 0 | 0 | 0 | 0 | 380,304 | 17,148,794 |

TABLE X
DIFFERENT RISK INDICES OF FUTURE TX-123BT WITH TEP AND GENERATION INVESTMENT

| Index | P1 | P2 | P3 | P4 | P5 | P6 |
|---|---|---|---|---|---|---|
| Annual LOLP | 0% | 0% | 0% | 0% | 0% | 0.0626% |
| LOLE (Hours/Bus) | 0 | 0 | 0 | 0 | 0 | 24.0 (0.29%) |
| EUE (MWh) | 0 | 0 | 0 | 0 | 0 | 419,486 |

The LOLE can indicate the expected total outage duration for a specific period, such as one year. In our LOLE calculation, we calculate the average outage hour on a bus for the entire year. Compared with LOLP, the LOLE can give us an insight on how long the load loss will last, instead of the occurrence probability of the load loss. As a brief conclusion, the LOLP, LOLE and EUE can comprehensively evaluate the load loss occurrence probability, duration, and the scale.

$$\text{LOLE} = (\sum_{p \in P, d \in D^{T(D)}, n \in B} N_{p,d,n}^{\text{H}} \cdot N^{\text{D}} + \sum_{p \in P, d \in D^{T(E)}, n \in B} N_{p,d,n}^{\text{H}} \cdot N^{\text{E}})/N^{\text{B}} \quad (34)$$

The results of calculated reliability indices based on the SCUC results for FR, FGI and FGTI cases, are shown in Tables VIII-X. With the transmission investments planned by TEP-CI, all three reliability indices for 2041-2045 decreased significantly. For the period 2046-2050, the LOLP and EUE both reduced significantly due to the transmission investments. LOLE has not decreased as substantially as the other two indicators, this suggests that while the severity of outages has been significantly mitigated, their durations remain prolonged. Based on the results and analysis, future grid reliability sees marked improvement with the transmission investments obtained by the TEP-CI strategy.

VI. CONCLUSION

As we expect more weather-dependent renewable resources in the future grids, improvement to the current TEP is required. A systematic procedure including data preparation, model improvement and reliability evaluation for the TEP-CI is presented in this paper. To address the weather and climate impact on the future power system that will be considered in the TEP-CI, the future weather-dependent spatio-temporal profiles for the TX-123BT test system are created. The improved TEP-CI model is proposed by considering these representative profiles in each planning epoch. The SCUC simulations are conducted on the future grids under different investment cases including FR, FGI, and FGTI. The reliability indices are proposed and calculated for each future planning epoch based on the daily operation conditions. The reliability of the grid under FR, FGI, and FGTI are compared and analyzed. This study depicts the organized scheme of the transmission planning considering weather and climate impact and paves the way for further planning studies.

ACKNOWLEDGMENT

This study is supported by the Alfred P. Sloan Foundation. We extend our gratitude to Dr. H. Li and his team for providing clean climate data, and to Dr. E. Yang and his team for providing ABM-based future generation investment data.

**Jin Lu** received the B.S. degree in electrical engineering from Dalian Maritime University, Dalian, Liaoning, China, in 2019, and M.S. degree in electrical engineering from the University of Houston, Houston, TX, USA, in 2020. He is currently pursuing the Ph.D. degree in electrical engineering at the University of Houston, Houston, TX, USA. From 2020 to 2024, he was a Research Assistant with the Renewable Power Grid Lab at the University of Houston. His research interests include power system operations and planning, power system restoration, renewable power systems with hydrogen energy.

**Xingpeng Li** received the B.S. degree in electrical engineering from Shandong University, Jinan, China, in 2010 and the M.S. degree in electrical engineering from Zhejiang University, Hangzhou, China, in 2013. He received the M.S. degree in industrial engineering in 2016 and the Ph.D. degree in electrical engineering in 2017 from Arizona State University, AZ, USA. He also received the M.S. degree in computer science from Georgia Institute of Technology, Atlanta, GA, USA, in 2023. From 2017 to 2018, he was a senior application engineer with ABB (now Hitachi Energy), San Jose, CA. Since 2018, he has been an Assistant Professor with the Department of Electrical and Computer Engineering at the University of Houston, TX, USA. His research interest includes the application of machine learning and optimization methods in power energy systems, power system control, operations and planning, grid integration of renewable generation, microgrids optimal sizing and energy management, distributed energy resources, energy storage, and battery degradation.